\documentclass[%
 reprint,
 amsmath,amssymb,
 aps,
]{revtex4-1}
\usepackage{siunitx}
\usepackage{graphicx}
\usepackage{dcolumn}
\usepackage{bm}
\usepackage{color}
\usepackage{verbatim}
\usepackage[T1]{fontenc}
\usepackage[titletoc]{appendix}
\usepackage{subfigure}
\usepackage{float}

\begin{document}

\preprint{APS/123-QED}

\title{Impact of short-range wakefields from radio-frequency cavity resonant modes on bunch lengthening}

\author{Tianlong He}\email{htlong@ustc.edu.cn}
\author{Jincheng Xiao}
\author{Weiwei Li}
\author{Zhenghe Bai}
\author{Weimin Li}

\affiliation{National Synchrotron Radiation Laboratory, University of Science and Technology of China, Hefei, Anhui, 230029, China}%

\date{\today}

\begin{abstract}
The high-performance operation of fourth-generation synchrotron light sources critically depends on harmonic cavities (HCs) to alleviate statistical collective effects through bunch lengthening. Active HCs are preferred over passive ones for achieving theoretically optimum bunch lengthening in Timing-mode operation, which is characterized by low average current and high bunch charge. This advantage stems from their ability to control cavity voltages via generator current. However, this study reveals a previously overlooked limitation: the detrimental impact of short-range wakefields from RF cavity resonant modes on bunch lengthening at high bunch charge. Using Hefei Advanced Light Facility parameters, we demonstrate that these wakefields can significantly degrade bunch lengthening. Further analysis with PETRA-IV parameters reveals that fine-tuning the HC voltage can optimize bunch lengthening. Notably, under specific HC settings, there exists two equilibrium bunch distributions. Our findings highlight the critical influence of RF cavity short-range wakefields on beam dynamics at high bunch charge, emphasizing their essential inclusion in the evaluation and optimization of HC performance for new-generation synchrotron light sources.

\begin{description}
	
\item[PACS numbers]
29.27.Bd, 41.75.Ht
\end{description}
\end{abstract}

\pacs{Valid PACS appear here}
\maketitle


\section{Introduction}
Bunch lengthening without an increase in bunch energy spread can effectively improve beam quality. This includes increasing the Touschek lifetime and mitigating the transverse emittance growth caused by intrabeam scattering. These improvements are crucial for ensuring the high-performance operation of fourth-generation synchrotron light sources~\cite{Nagaoka01}. The most effective measure for bunch lengthening is the use of an additional harmonic cavity (HC) system. This system works in conjunction with the main cavity (MC) system to widen the potential well, thereby stretching the bunch length. When the optimum lengthening condition is achieved—specifically, when the first and second derivatives of the total voltage of the double radio-frequency (RF) systems are equal to zero~\cite{Byrd02}—the bunch length can be increased to at least four times its natural length. This level of bunch lengthening can significantly enhance the overall performance of synchrotron light sources and is commonly desired~\cite{Cullinan03, He04, Tavares05, Carmignani06, Schroer07, Jiao08, APSU09}.

Bunch lengthening can be limited by coupled bunch instabilities, such as mode 1 and mode 0 instabilities driven by the fundamental mode impedance of HCs~\cite{He11, Marco12, He13, He14}. One special mode 1 instability, also known as periodic transient beam loading instability, has garnered significant attention in recent studies due to its potential to limit bunch lengthening at relatively high beam currents~\cite{Yamamoto15, Gamelin16, Bellafont17}. Additionally, parasitic modes that are not well-damped in RF cavities can cause coupled bunch instabilities, characterized by oscillations in bunch length and increased bunch energy spread~\cite{Cullinan18}. To mitigate these instabilities, fine temperature tuning is often employed to shift the frequencies of parasitic modes away from the beam spectrum lines~\cite{Tavares19}.
For both fundamental and parasitic modes, the primary concern is their long-range dynamic effects related to coupled bunch instabilities, whose growth rates are proportional to the beam current. 

Some synchrotron light sources need operate in timing mode (e.g., few-bunch mode), where a low average current is achieved by reducing the number of electron bunches, while maintaining high bunch charge to deliver intense, time-resolved X-ray pulses~\cite{Olsson}. For timing mode, active HCs are preferred over passive ones for bunch lengthening. This preference arises because, at low currents, passive HCs cannot achieve the desired voltage levels, whereas active HCs can be regulated via generator current to meet the required voltage and phase condition for optimum bunch lengthening. In previous analyses of active HC bunch lengthening, ideal cavity voltage formulas were generally directly used for analytical calculations of bunch lengthening~\cite{Haishen, Liu, Bassi}. While these analytical calculations are generally accurate for low bunch charge, they exhibits significant errors at high bunch charge. 

For high-bunch-charge scenarios, the focus has traditionally been on the short-range wakefield effects from the broadband impedance of the entire ring, such as potential well distortion, bunch elongation (in rings with positive momentum compaction factors), and microwave instabilities~\cite{Chao10}. However, the short-range wakefield effects from RF cavity resonant modes have been rarely discussed. The strength of these wakefields generally scales with $R/Q\times\omega_r\times q$, where $R/Q$ is the shunt impedance divided by the quality factor, $q$ denotes the bunch charge, and $\omega_r$ is the resonant frequency. The fundamental mode has the highest $R/Q$ but the lowest frequency, while parasitic modes typically have smaller $R/Q$ values but higher frequencies. As a result, when multiple RF cavities are present, their combined short-range effects on bunch lengthening cannot be ignored, especially at high bunch charges.

In this paper, we investigate the influence of short-range effects from RF cavity resonant modes on bunch lengthening, focusing on high-bunch-charge filling scenarios. To facilitate this investigation, we assume a uniform filling pattern and equilibrium conditions, which allows us to iteratively calculate the single-bunch equilibrium distribution. The remainder of this paper is organized as follows. In Sec.~\ref{sec:level2}, we introduce the analytical method and distinguish between the long-range and short-range effects from cavity resonant modes. In Sec.~\ref{sec:level3}, we use the Hefei Advanced Light Facility (HALF) storage ring as an example to demonstrate the influence of the short-range effects of resonant modes in RF cavities on bunch lengthening under the well-known optimum lengthening condition. In Sec.~\ref{sec:level4}, we analyze the influence of the short-range effects on bunch lengthening for the PETRA-IV storage ring. Finally, conclusions are presented in Sec.~\ref{sec:level5}.

\section{\label{sec:level2} Theory and method}
In this paper, we assume that the double RF systems use the fundamental mode, i.e., TM010, as the working mode, and we mainly focus on analyzing the impact of short-range wakefields of the fundamental mode on bunch lengthening. For the completeness of theoretical expression, we will first review the theoretical framework of beam-cavity interaction. For convenience in subsequent discussions, we will distinguish between the short-range and long-range effects of beam-fundamental mode interaction. The short-range term represents the immediate electromagnetic interaction between the bunch current and fundamental mode, while the long-range term corresponds to the steady-state beam loading voltage under steady-state assumptions.

\subsection{\label{sec:level2.1} Steady-state cavity voltage phasors}
\begin{figure}[!htbp]
  \centering
     \includegraphics[width=0.38\textwidth]{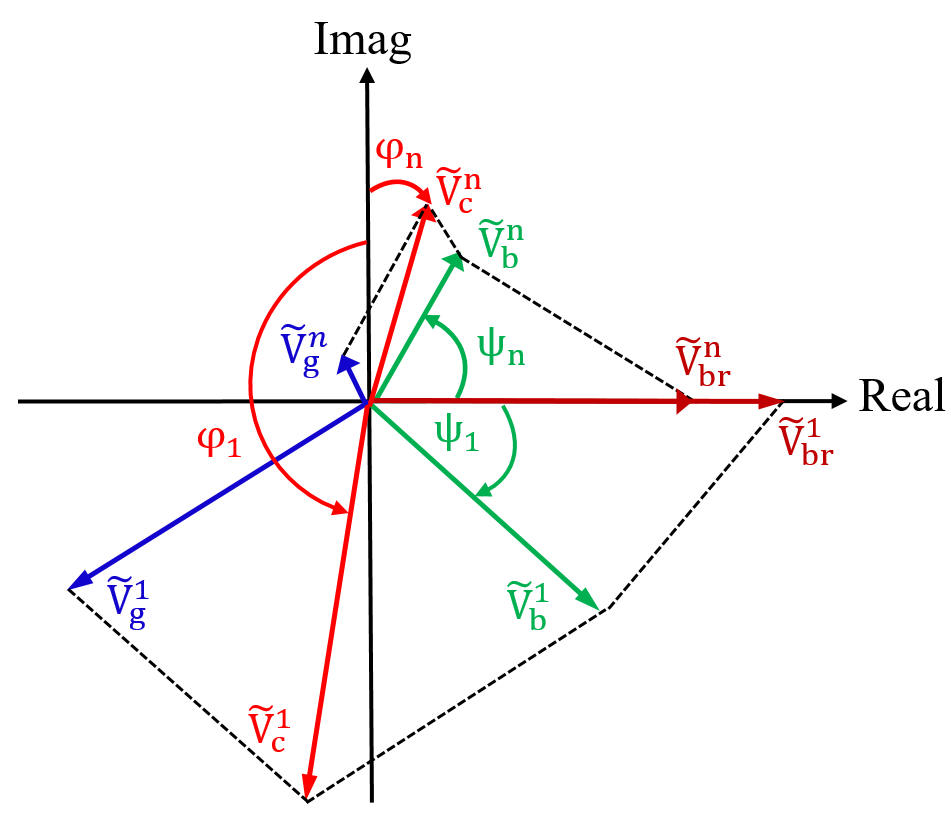}
  \caption{Voltage phasors of the double active RF cavities. $\widetilde{V}_{b}$, $\widetilde{V}_{g}$, and $\widetilde{V}_{c}$ denote the beam loading, generator, and total cavity voltage phasors, respectively. $\widetilde{V}_{br}$ represents the resonant beam loading voltage phasors. The symbol $\psi$ denotes the detuning phase, while $\varphi$ signifies the synchronous phase. The subscripts (or superscripts) $1$ and $n$ correspond to the MC and HC, respectively.}
  \label{fig1}
\end{figure}

Under the steady-state assumption, the long-range interaction can be represented as a steady-state cavity voltage phasor~\cite{Ng20}. For simplicity, we represent the cavity voltages as phasors that rotate counterclockwise, and in the complex plane, take the real-axis positive direction as the beam image current direction. Due to the MC and HC resonant frequencies being less and larger than the harmonics of beam revolution frequency, the MC and HC beam-loading voltage phasors will be located in the fourth and first quadrants, respectively, as shown in Fig.~\ref{fig1}. The angle formed by the beam-loading voltage phasor and the beam image current phasor is called the detuning phase, which are 0-90 degrees for HC and -90-0 degrees for MC.

The resonant modes in RF cavities are in general described as the R-L-C parallel circuit, which have the same impedance form as
\begin{equation}\label{eq:labe1}
   Z(\omega)=\frac{R}{1+jQ(\frac{\omega_r}{\omega}-\frac{\omega}{\omega_r})},
\end{equation}
where $R$, $Q$, $\omega_r$ are the characteristic parameters of the resonant mode, representing the shunt impedance, quality factor, and angular resonant frequency, respectively. Note that the shunt impedance and quality factor should be considered as loaded values if the cavity has a power coupler.

For the fundamental mode, under uniform filling patterns, the beam-induced stationary voltage phasor is well-known as
\begin{equation}\label{eq:labe2}
   \widetilde{V}_{b}=2\widetilde{F}I_0 R \cos{(\psi)} e^{j\psi},
\end{equation}
where $\widetilde{F}$ is the complex bunch form factor, $I_0$ is the average beam current, and $\psi$ is the detuning phase, which can be computed by
\begin{equation}\label{eq:labe3}
   \tan{(\psi)}=\frac{2Q\Delta \omega}{\omega_r},
\end{equation}
where $\Delta \omega=\omega_r-i\omega_{rf}$ is the angular detuning frequency, $i=1$ and $n$ correspond to the MC and HC, respectively, $n$ is the harmonic number of HC, and $\omega_{rf}$ is the angular RF frequency.
For active RF cavities, the total cavity voltage phasor is the vector sum of generator and beam induced voltage phasors:
\begin{equation}\label{eq:labe4}   \widetilde{V}_{c}=\widetilde{V}_{g}+\widetilde{V}_{b}
\end{equation}
Normally, the generator voltage can be adjusted to make the total voltage equal to the target set voltage.

The total cavity voltage phasor, once determined, can be directly expressed in sine-convention as:
\begin{equation}\label{eq:labe5}
   V^{1}_c(\varphi)=V_{1} \sin{(\varphi+\varphi_{1})},
\end{equation}
\begin{equation}\label{eq:labe6}
   V^{n}_c(\varphi)=V_{n} \sin{(n\varphi+\varphi_{n})},
\end{equation}
where $V_{1,n}$ and $\varphi_{1,n}$ are the amplitude and reference phase, respectively.

Next, it will be proven that Eqs.~(\ref{eq:labe5}) and~(\ref{eq:labe6}) only include the long-range effect of beam-fundamental mode interaction.

\subsection{\label{sec:level2.2} Long-range effect}
Assuming that the ring has $h$ buckets which are filled with $m$ equally spaced and equally charged bunches ($h/m$ is an integer). We further assume that these bunches have identical normalized equilibrium distribution of $\rho(\tau)$. Let us first deduce the cavity voltage induced by a single passage of one bunch through the cavity. As is well known that this voltage is equivalent to~\cite{He21}
\begin{equation}\label{eq:labe7}
   \widetilde{V}_{0}=\frac{q\omega_r R}{Q}\int_{-\infty}^{\infty}\rho(\tau) e^{(-j+\frac{1}{2Q})\omega_r \tau}\,d\tau,
\end{equation}
where $q$ is the bunch charge. The integral part is simply denoted by $\widetilde{F}=\int_{-\infty}^{\infty}\rho(\tau) e^{(-j+\frac{1}{2Q})\omega_r \tau}\,d\tau$, being the complex bunch form factor. Then Eq.~(\ref{eq:labe7}) can be simplified to
\begin{equation}\label{eq:labe8}
   \widetilde{V}_{0}=\frac{\widetilde{F} q\omega_r R}{Q}.
\end{equation}

After the Nth passage of the bunches, the beam-induced cavity voltage can be expressed as
\begin{equation}\label{eq:labe9}
   \widetilde{V}_{N}=\widetilde{F}\frac{q\omega_r R}{Q}\sum\limits^{N}_{i=0} e^{(j-1/2Q)\omega_r T_0 i/m},
\end{equation}
where $T_0$ is the revolution time. Note that $\widetilde{V}_{N}$ will converge to a certain value as $N$ approaches infinity. With $\alpha=(j-1/2Q)\omega_r T_0$, Eq.~(\ref{eq:labe9}) can be simplified as
\begin{equation}\label{eq:labe10}
   \widetilde{V}_{N}=\frac{\widetilde{F} q\omega_r R/Q}{1-e^{\alpha/m}}.
\end{equation}

As previously mentioned, the long-range effect refers to the cavity voltage excited and accumulated by the previous bunch of passages through RF cavities. For a certain bunch passing through RF cavities, the steady-state cavity voltage as seen by this bunch undergoes a period of evolution (amplitude decaying and phase rotation) compared to when the previous bunch passed through. This period is equal to the interval of $T_0/m$ between adjacent bunches. Consequently, the cavity voltage phasor observed by the current bunch becomes

\begin{equation}\label{eq:labe11}
   \widetilde{V}_{L}=\frac{\widetilde{F} q\omega_r R/Q e^{\alpha/m}}{1-e^{\alpha/m}}=\frac{\widetilde{F} q\omega_r R/Q}{e^{-\alpha/m}-1}.
\end{equation}
When m is much larger than 1, we have
\begin{equation}\label{eq:labe12}
   \begin{aligned}
   e^{-\alpha/m} &=e^{-j\omega_r T_0/m} e^{\frac{\omega_r T_0}{2Qm}}\\
   &=e^{-j\Delta\omega T_0/m} e^{\frac{\omega_r T_0}{2Qm}}\\
   &\approx 1-j\Delta\omega T_0/m+\frac{\omega_r T_0}{2Qm}.
   \end{aligned}
\end{equation}
Substituting Eq.~(\ref{eq:labe12}) into Eq.~(\ref{eq:labe11}), it gives
\begin{equation}\label{eq:labe13}
   \begin{aligned}
   \widetilde{V}_{L} &\approx \frac{\widetilde{F} mq\omega_r R/Q}{\frac{\omega_r T_0}{2Q}-j\Delta\omega T_0}\\
   &=\frac{2\widetilde{F} R mq/T_0}{1-j2Q\Delta\omega/\omega_r}\\
   &=2\widetilde{F}I_0 R \cos{(\psi)} e^{j\psi}.
   \end{aligned}
\end{equation}
where $\psi=\arctan(\frac{2Q\Delta\omega}{\omega_r})$, being the detuning phase. It should be noted that when $m$ approaches 1, the approximation taken in Eq.~(\ref{eq:labe12}) will introduce a certain error in Eq.~(\ref{eq:labe13}), and Eq.~(\ref{eq:labe10}) should be used to calculate the steady-state beam loading voltage phasor. Note also that Eq.~(\ref{eq:labe13}) and Eq.~(\ref{eq:labe2}) are exactly the same, indicating that Eq.~(\ref{eq:labe2}) only represents the long-range effect of beam-fundamental mode interaction.

\subsection{\label{sec:level2.3} Short-range effect}
The short-range effect can be described directly with the corresponding Green’s wakefield function of resonant modes~\cite{Palumbo22}
\begin{equation}\label{eq:labe14}
   W_{SR}(\tau)=\frac{\omega_r R}{Q} e^{-\frac{\omega_r \tau}{2Q}} [\cos(\omega_n \tau)-\frac{\omega_r}{2Q\omega_n}\sin(\omega_n \tau)] H(\tau),
\end{equation}
where $H(\tau)$ is the Heaviside step function, and $\omega_n$ is written as
\begin{equation}\label{eq:labe15}
   \omega_n=\frac{\omega_r \sqrt{4Q^2-1}}{2Q}.
\end{equation}
For a narrowband resonator with large $Q$ value, its wake function can be simplified as
\begin{equation}\label{eq:labe16}
   W_{SR}(\tau)=\frac{\omega_r R}{Q} e^{-\frac{\omega_r \tau}{2Q}} \cos(\omega_r \tau) H(\tau).
\end{equation}
Note that Eq.~(\ref{eq:labe16}) can be used to describe the beam-cavity resonant mode interaction for one bunch at the present passage through RF cavities. 

It is clear that the short-range wakefield strength fully depends on the product of $R/Q$ and $\omega_r$, indicating that those resonant modes having high $R/Q\times\omega_r$ possibly lead to a strong effect.

\subsection{\label{sec:level2.4} Beam equilibrium distribution}
We emphasize here that the long-range and the short-range effects of beam-cavity resonant mode interaction have been distinguished as mentioned above. For those light sources utilizing the double active RF systems, the steady-state total RF voltage corresponding only to the long-range effect can be given as
\begin{equation}\label{eq:labe17}
   V_T(\tau) = V_{1}\sin(\omega_{rf}\tau+\varphi_1)+ k V_{1}\sin(n\omega_{rf}\tau+\varphi_n),
\end{equation}
where $k=V_1/V_n$ is the voltage ratio of HC to MC.

The optimal lengthening condition can be reached by setting the first and second derivatives of the total RF voltage to zero. Under this condition, the RF potential is flat at the center, hence also known as the flat-potential condition (FP). To keep it fulfilled, we have~\cite{Byrd02}
\begin{equation}\label{eq:labe18}
    k_{fp}=\sqrt{\frac{1}{n^2}-\frac{1}{n^2-1}\Big{(}\frac{U_0}{eV_{1}}\Big{)}^2},
\end{equation}
\begin{equation}\label{eq:labe19}
    \tan{(\varphi_{n,fp})}=-\frac{n U_0}{eV_{1}}\frac{1}{\sqrt{(n^2-1)^2-\Big{(}\frac{n^2 U_0}{eV_{1}}\Big{)}^2}},
\end{equation}
\begin{equation}\label{eq:labe20}
    \sin{(\varphi_{1,fp})}=\frac{n^2}{n^2-1}\frac{U_0}{eV_{1}},
\end{equation}
where $e$ denotes the elementary electron charge, $\varphi_{1,fp}$ and $\varphi_{n,fp}$ are the flat-potential synchronous phases of the MC and HC, respectively, and $k_{fp}$ is the flat-potential voltage ratio of HC to MC. Note that the above three parameters are obtained in the absence of the short-range effects. 

The RF potential can be written as
\begin{equation}\label{eq:labe21}
   \begin{aligned}
   \Phi_{RF}(\tau) = \frac{eV_{1}}{2\pi hE_0\alpha_c\sigma_{\delta}^2}\{\cos(\varphi_1)-\cos(\omega_{rf}\tau+\varphi_1)+\\
   \frac{k}{n}[\cos(\varphi_n)-\cos(n\omega_{rf}\tau+\varphi_n)]
    -\frac{U_0\omega_{rf}\tau}{eV_{1}}\},
    \end{aligned}
\end{equation}
where $h$ is the harmonic number, $E_0$ is the beam energy, $\alpha_c$ is the momentum compaction factor, $\sigma_\delta$ is the rms energy spread, and $U_0$ is the energy loss per turn.

The short-range potential can be given as
\begin{equation}\label{eq:labe22}
   \Phi_{SR}(\tau)=\frac{q}{E_0T_0\alpha_c\sigma_{\delta}^2}\int_{0}^{\tau}d\tau^{\prime}\int_{-\infty}^{\infty}\rho(\tau^{\prime\prime})W_{SR}(\tau^{\prime}-\tau^{\prime\prime})\,d\tau^{\prime\prime},
\end{equation}
where $W_{SR}(\tau)$ can be treated as the superposition of the different short-range potentials if considering multiple resonant modes.

The total longitudinal potential is the sum of the long- and short-range potentials
\begin{equation}\label{eq:labe23}
   \Phi_{T}(\tau)=\Phi_{RF}(\tau)+\Phi_{SR}(\tau).
\end{equation}
Then the bunch equilibrium density distribution can be computed with
\begin{equation}\label{eq:labe24}
  \rho(\tau)=\frac{e^{\Phi_T(\tau)}}{\int_{-\infty}^{\infty} e^{\Phi_T(\tau)}\,d\tau}.
\end{equation}
Note that Eqs.~(\ref{eq:labe21})-(\ref{eq:labe24}) exactly form the well-known Ha{\"\i}ssinski equation~\cite{Haissinski23}, and the equilibrium distribution requires iterative solution. 

In the following sections, we will focus on the optimum bunch lengthening condition for the double active RF systems. Under this condition, the double RF cavity voltages can be determined using Eqs.~(\ref{eq:labe18})-(\ref{eq:labe20}). As a consequence, the HC voltage ratio and phase can be plugged directly into the Ha{\"\i}ssinski equation to iteratively calculate the equilibrium distribution. This solution is simple and not time-consuming, after all, in case of the uniform filling, only a single bunch needs to be calculated. While in case of the arbitrary filling, there are several existing self-consistent algorithms available~\cite{He21,Warnock24}.

As indicated in Eq.~(\ref{eq:labe22}), the short-range potential strength is proportional to bunch charge and short-range wakefield strength. Consequently, in case of synchrotron light storage rings operating in timing mode with high bunch charge and employing multiple RF cavities, the short-range potential can substantially influence the equilibrium bunch distribution.

\section{\label{sec:level3} Influence of the short-range effect on bunch lengthening}
The HALF storage ring~\cite{Bai25} is taken as a study case to investigate the influence of the short-range effect on bunch lengthening. Table~\ref{table1} shows the main parameters of HALF, including the characteristic parameters of the double RF cavities and the corresponding RF voltages and phases under the well-known flat-potential condition. It should be noted that both the MC and HC in HALF are superconducting, with the HC operating in a passive mode. For analytical simplicity, we adopt the assumption of an actively powered HC in this study.
\begin{table}[!hbt]
\setlength{\tabcolsep}{3.0mm}
   \centering
   \caption{Main parameters of HALF.}
   \begin{tabular}{lc}
       \toprule
       \textbf{Parameter} &{Value} \\
       \hline
Energy &~~~~~2.2 GeV       \\
Circumference &~~~~~479.86 m \\
Current &~~~~~350 mA \\
Momentum compaction &~~~~~$9.4\times10^{-5}$ \\
Harmonic number &~~~~~800 \\
Radiation energy loss per turn &~~~~~400 keV \\
RMS energy spread &~~~~~$7.44\times10^{-4}$ \\
Bunch charge (completely filling) &~~~~~0.7 nC \\
499.8 MHz MC $R/Q$ &~~~~~45 $\Omega$ \\ 
499.8 MHz MC loaded $Q$ &~~~~~$1\times10^{5}$ \\ 
3rd HC $R/Q$ &~~~~~39 $\Omega$ \\ 
3rd HC loaded $Q$ &~~~~~$2\times10^{8}$ \\ 
MC voltage amplitude &~~~~~1200 kV \\ 
MC voltage phase (FP)  &~~~~~158.0 deg \\ 
HC voltage amplitude (FP) &~~~~~374.2 kV \\ 
HC voltage phase (FP)  &~~~~~-7.68 deg\\
       \hline
   \end{tabular}
   \label{table1}
\end{table} 

\subsection{\label{sec:level3.1} Short-range effects of fundamental mode}

\begin{figure}[!htbp]  
  \centering
      \includegraphics[width=0.34\textwidth]{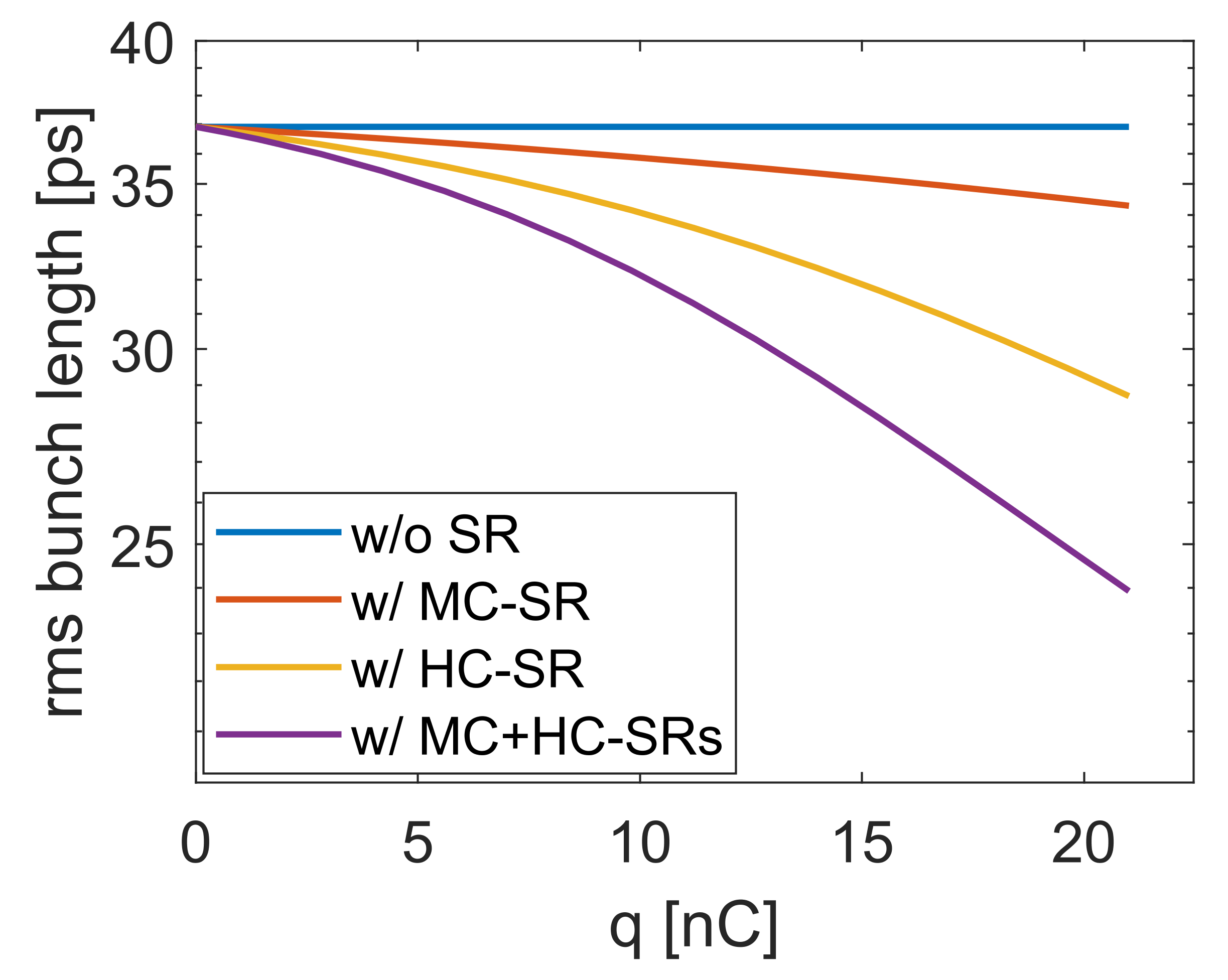}
  \caption{The rms bunch length as a function of the bunch charge.}
  \label{fig2}
\end{figure}

Four distinct cases are analyzed: (i)  the short-range potentials (SRs) are entirely ignored; (ii) only the MC SR is accounted for; (iii) only the HC SR is considered; and (iv) both SRs from the double RF cavities are incorporated. As shown in Fig.~\ref{fig2}, the inclusion of SRs from the double RF cavities consistently degrades the bunch lengthening as the bunch charge increases. Furthermore, the influence of the HC SR is found to be more significant than that of the MC SR. This indicates that the HC has a stronger short-range potential compared to the MC, which can be attributed to its frequency being three times that of the MC.

\begin{figure}[!htbp]  
  \centering
  \includegraphics[width=0.34\textwidth]{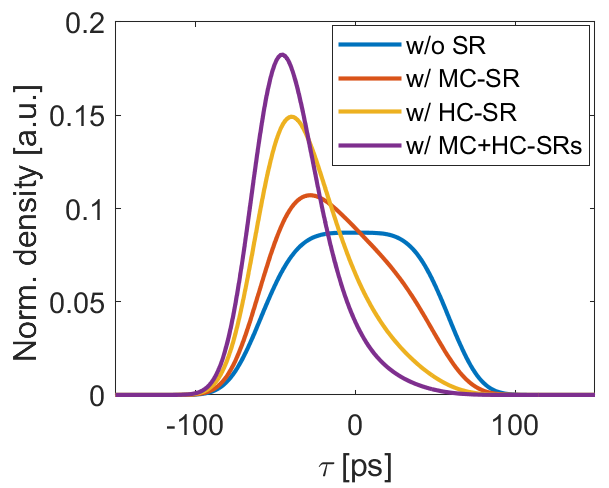}
  \caption{Bunch profiles for the case of assuming a high bunch charge of 20 nC.}
  \label{fig3}
\end{figure}

Figure~\ref{fig3} shows the corresponding bunch equilibrium profiles for a high bunch charge of 20 nC. Under such a high-bunch-charge scenario, the SRs of the double RF cavities are strong enough to cause the flat potential well distortion, making the bunch distribution no longer symmetrical and lean forward associated with reducing the bunch length. It should be pointed out that for HALF with at least 80 percent filling rate, the actual bunch charge is generally less than 1 nC, so it can be expected that the influence of the SPs from the RF cavities on bunch equilibrium profiles can be safely ignored.

\begin{figure}[!htbp]  
  \centering
  \includegraphics[width=0.34\textwidth]{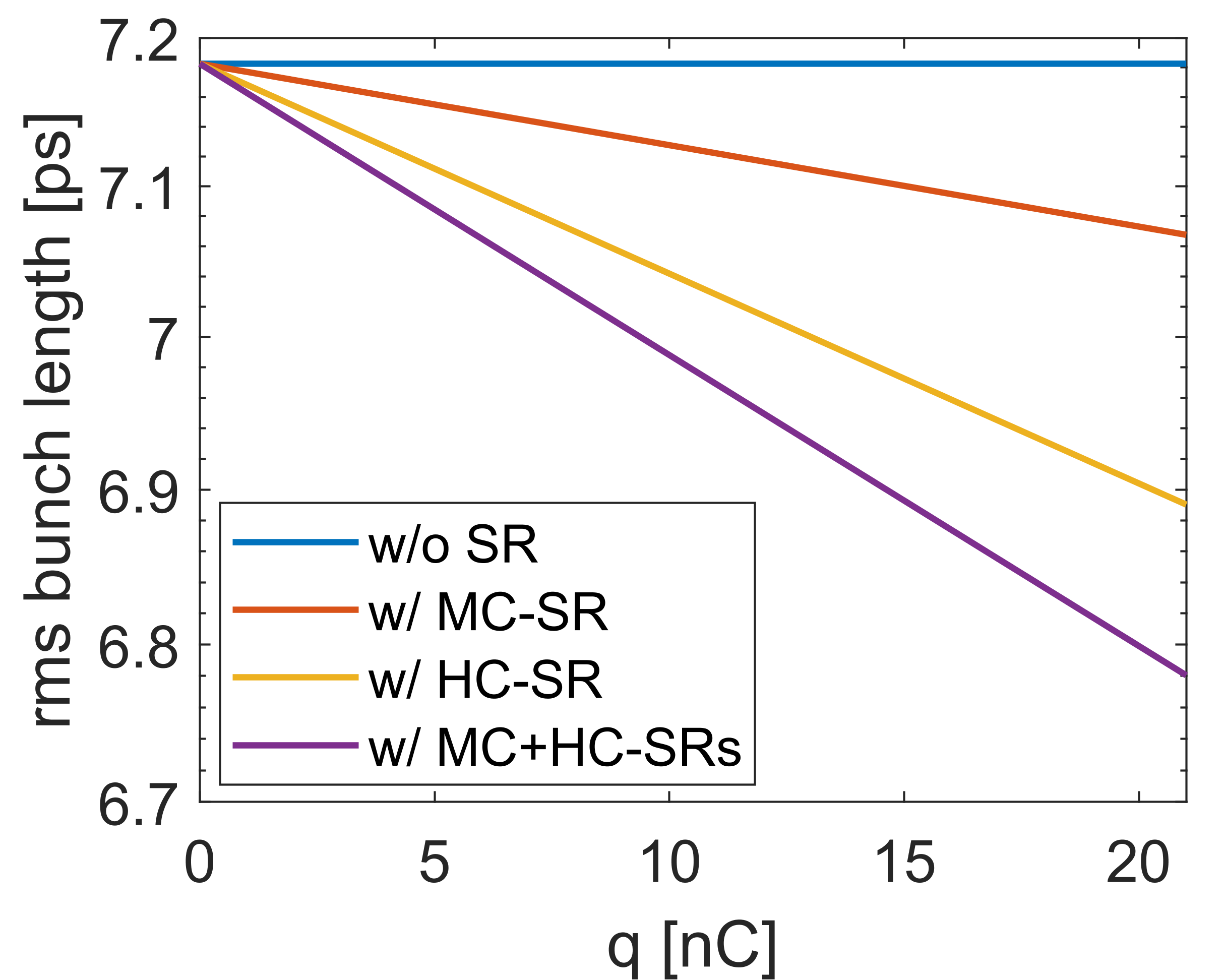}
  \caption{The rms bunch length as a function of the bunch charge.}
  \label{fig4}
\end{figure}

For the case of parking the HC (that is, minimizing its voltage), the resulting bunch length as a function of the bunch charge is shown in Fig.~\ref{fig4}. The same four cases as aforementioned are considered. It can be seen that the SRs also reduce the bunch length as increasing the bunch charge. Nevertheless, the reduction in bunch length is not so critical with respect to those shown in Fig.~\ref{fig2}.

In the above analysis, we see that the short-range effects have a relatively small impact on the bunch lengthening of HALF, which is attributed to the use of superconducting cavities with small $R/Q$ values of the fundamental mode. Now let us assume a scenario where the double RF system uses the normal-conducting cavities. We can select 500 MHz HOM BESSY-type damped cavity~\cite{Marhauser} and 1499 MHz HOM-damped ALBA cavity~\cite{Perez} as for MC and HC, respectively. Table~\ref{table2} shows the required number of both cavities, as well as the total R/Q values. 

\begin{figure}[!htbp]  
  \centering
      \includegraphics[width=0.35\textwidth]{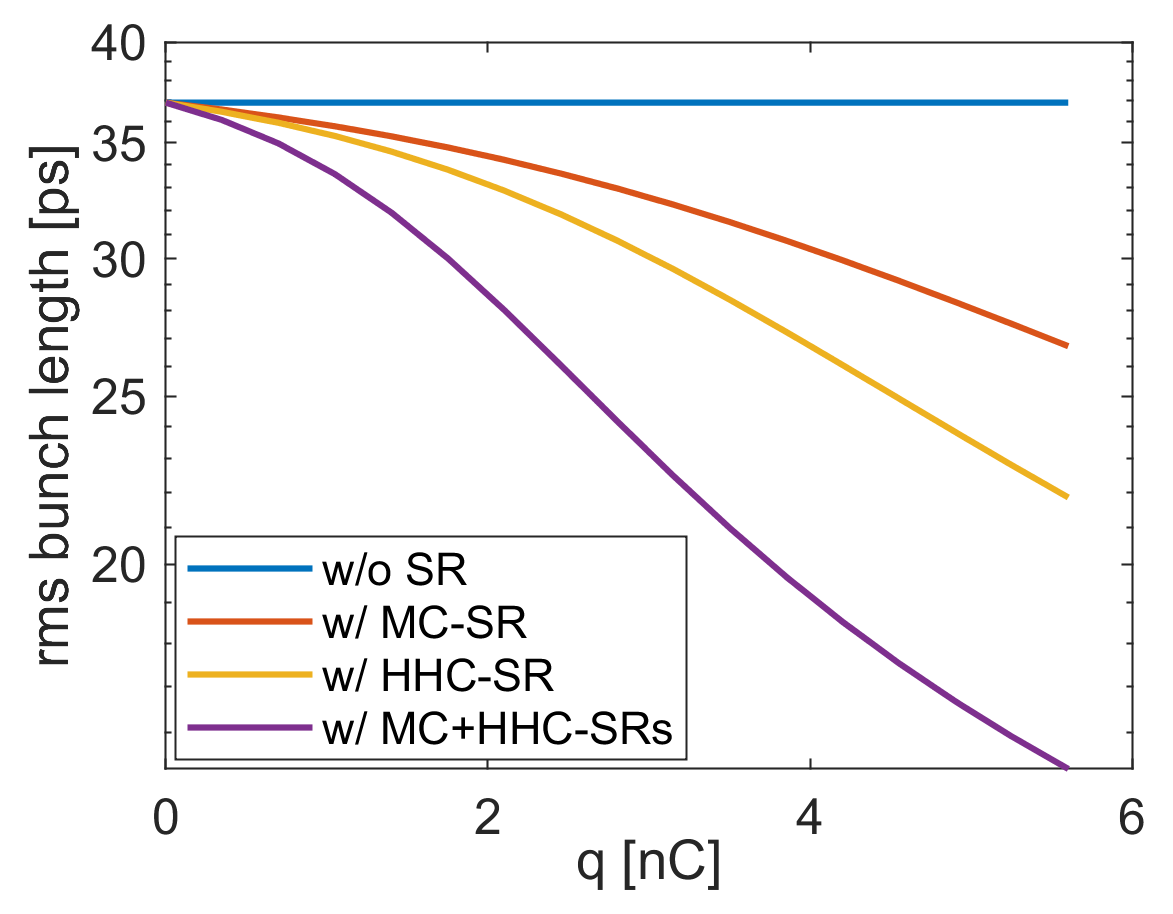}
  \caption{The rms bunch length as a function of the bunch charge.}
  \label{fig5}
\end{figure}

\begin{table}[!hbt]
\setlength{\tabcolsep}{3.0mm}
   \centering
   \caption{Bunch charge and cavity parameters of four high-energy synchrotron light sources.}
   \begin{tabular}{lccc}
       \toprule
       \textbf{Cavity} &{single $R/Q$} &{Cavity No.}  &{Total $R/Q$ } \\
       \hline
           MC & 115 $\Omega$& 4 & 460 $\Omega$\\
           HC & 80 $\Omega$& 3 & 240 $\Omega$\\
       \hline
   \end{tabular}
   \label{table2}
\end{table} 
Figure~\ref{fig5} shows the resulting bunch length including the four cases same as considered in Fig.~\ref{fig2}. When the bunch charge is larger than 4 nC, the short-range effects from cavity fundamental modes can degrade the bunch lengthening by half. For a timing mode designed for HALF, the bunch charge would be larger than 4 nC. Therefore, considering the influence of short-range effects on bunch lengthening, for the HALF project, the superconducting cavity scheme demonstrates superior advantages over the normal-conducting cavity approach. 

\subsection{\label{sec:level3.2} Discussion on other impact factors}
As indicated in Eqs.~\ref{eq:labe16} and~\ref{eq:labe22}, in addition to the bunch charge and $R/Q$, the short-range potential is also related to the quality factor and resonant frequency, whose influences on bunch lengthening will be discussed separately as follows.
\subsubsection{\label{sec:level3.2.1} Quality factor}
The quality factor ($Q$) of resonant modes in RF cavities can vary greatly. In the process of cavity design, the fundamental mode $Q$ is expected to be as large as possible, while the parasitic mode $Q$ is suppressed to be as low as possible through HOM damping technology. For active cavities in operation, the fundamental mode $Q$ may undergo significant changes due to different coupling factor settings. 

\begin{figure}[!htbp]  
  \centering
      \includegraphics[width=0.34\textwidth]{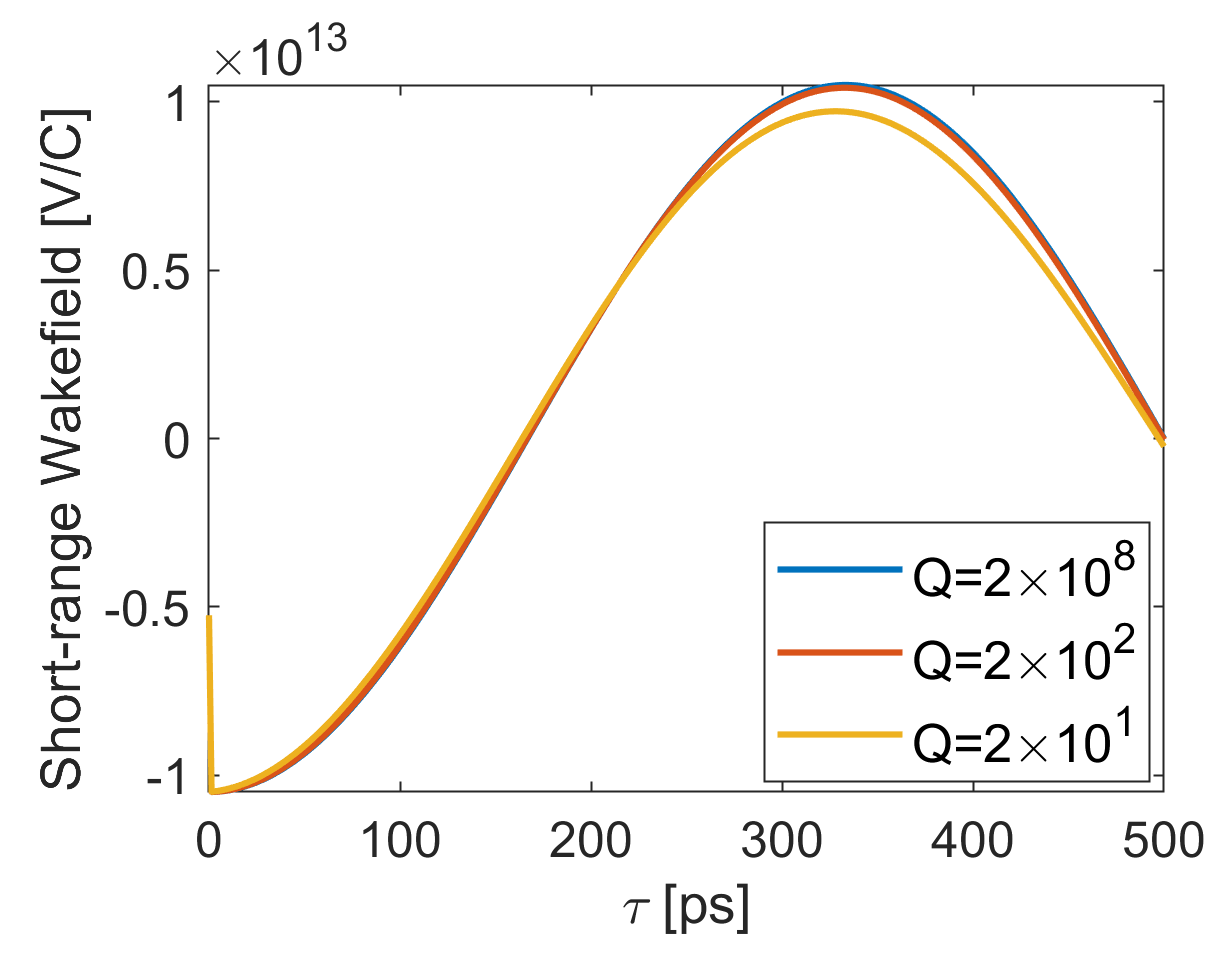}
  \caption{The short-range wakefield of the HC fundamental mode with different Q values.}
  \label{fig6}
\end{figure}

It then raises a question: will the variation in $Q$ value affect the short-range effects? Figure~\ref{fig6} shows the wakefield functions in a short range for the case of three different $Q$ values of HC fundamental mode. It can be seen that even when the $Q$ value is reduced to 20, it affects the short-range wakefield very slightly.

\begin{figure}[!htbp]  
  \centering
      \includegraphics[width=0.35\textwidth]{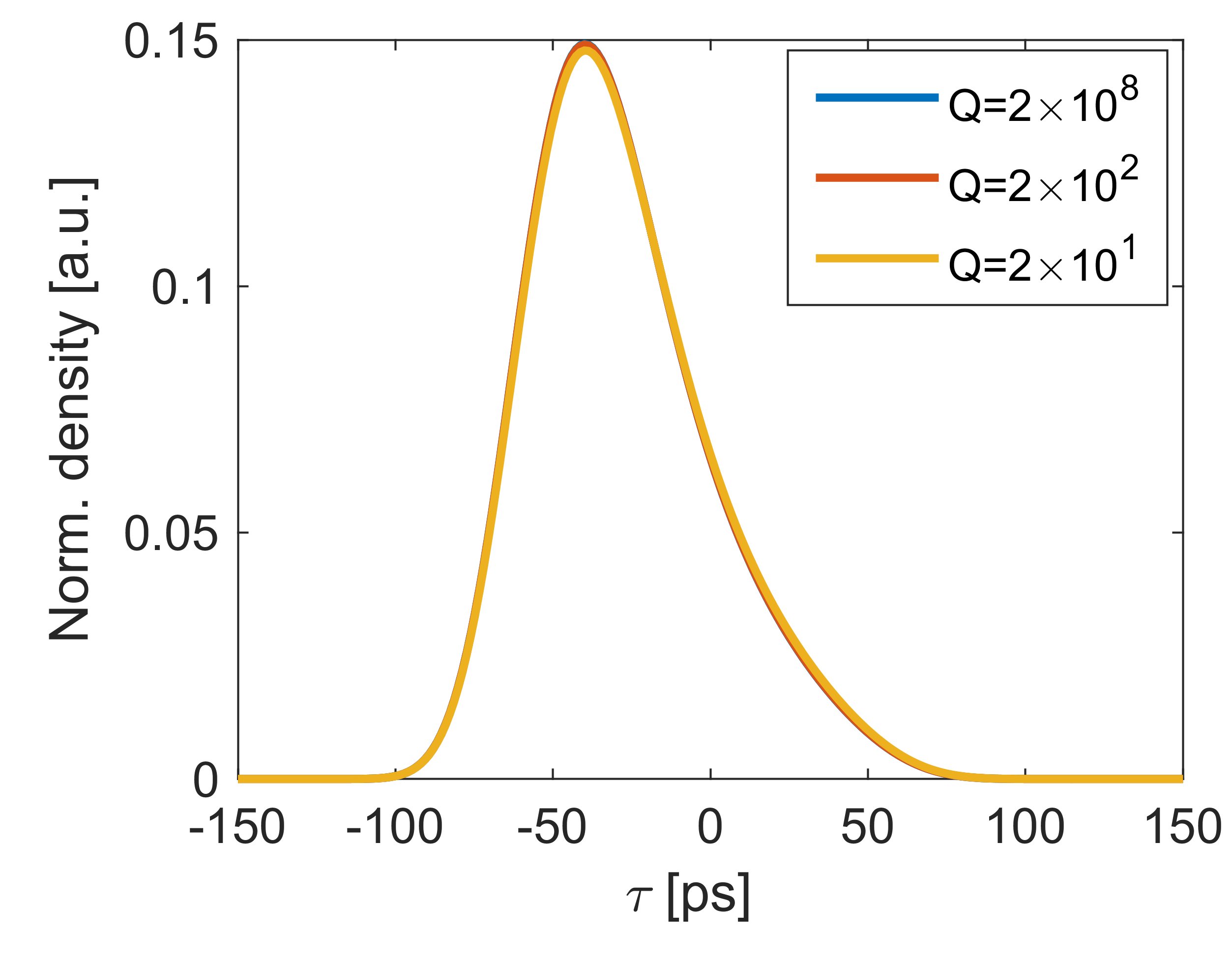}
  \caption{Bunch profiles for the case of assuming a high bunch charge of 20 nC.}
  \label{fig7}
\end{figure}

Figure~\ref{fig7} shows the corresponding bunch equilibrium distribution for an assumed case of high bunch charge of 20 nC. We can conclude that the variation of Q value in a very large range has almost no effects on the bunch equilibrium distribution.

\subsubsection{\label{sec:level3.2.2} Resonant frequency}
It has shown the influence of the short-range effects from the 500 MHz MC and 1499 MHz HC on bunch lengthening. Comparative analysis reveals that the HC exhibits a more pronounced influence on bunch lengthening owing to its higher resonant frequency relative to the MC. Considering the abundant parasitic modes in RF cavities, their short-range effects can also affect the bunch lengthening significantly when their $R/Q$ values are not negligible. Therefore, it is necessary to assess the influence of resonant frequencies. 

\begin{figure}[!htbp]  
  \centering  \includegraphics[width=0.36\textwidth]{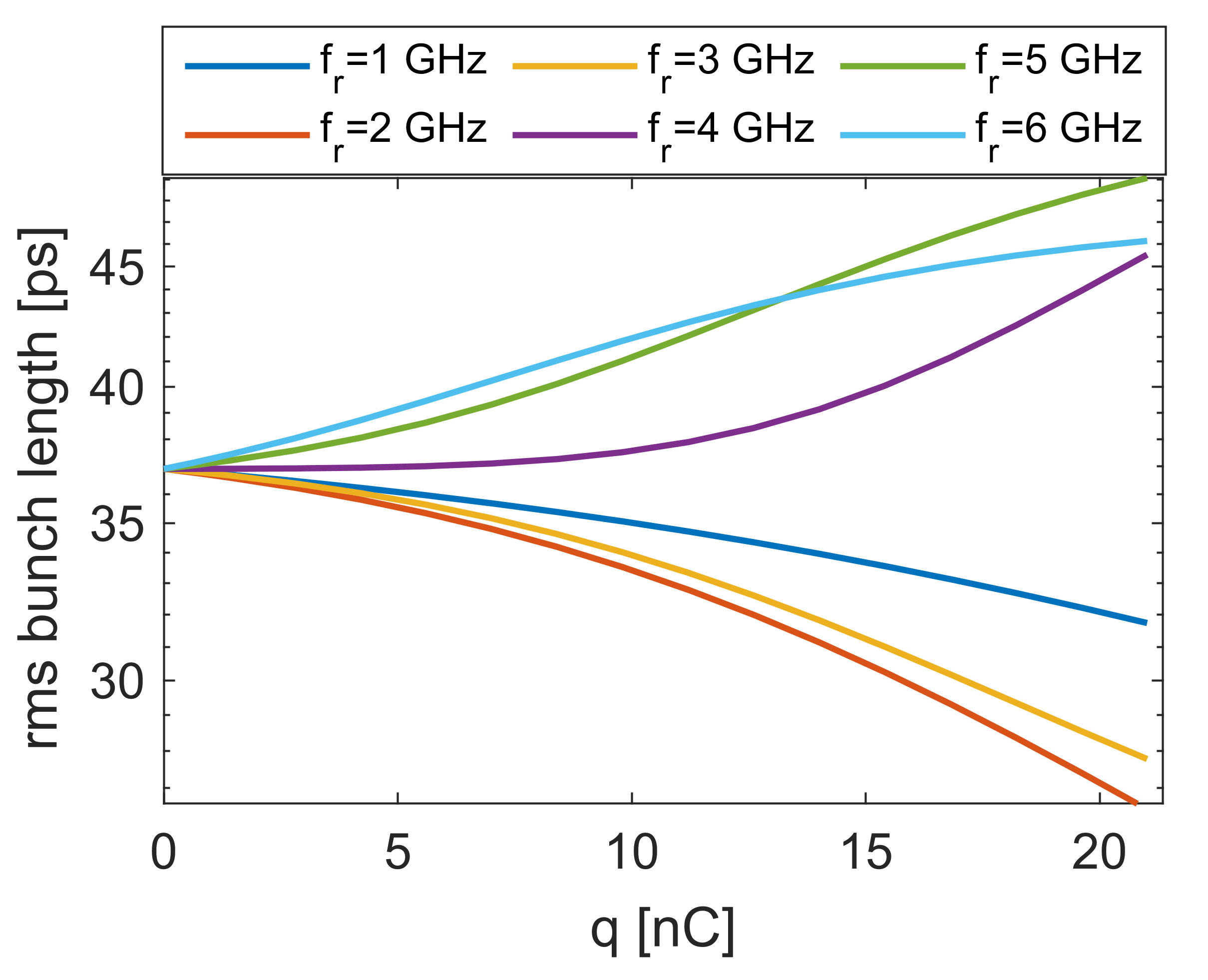}
  \caption{The rms bunch length as a function of the bunch charge. The legend gives the resonant frequency.}
  \label{fig8}
\end{figure}

Here we assume a single resonant mode with $R/Q$ of 39 $\Omega$ and $Q$ of $2\times10^8$, which are the same as those of the HC of HALF. The cases of six frequencies uniformly spaced in the range of 1$\sim$6 GHz are investigated. The resulting bunch length versus the bunch charge is shown in Fig.~\ref{fig8}.

\begin{figure}[!htbp]  
  \centering
      \includegraphics[width=0.35\textwidth]{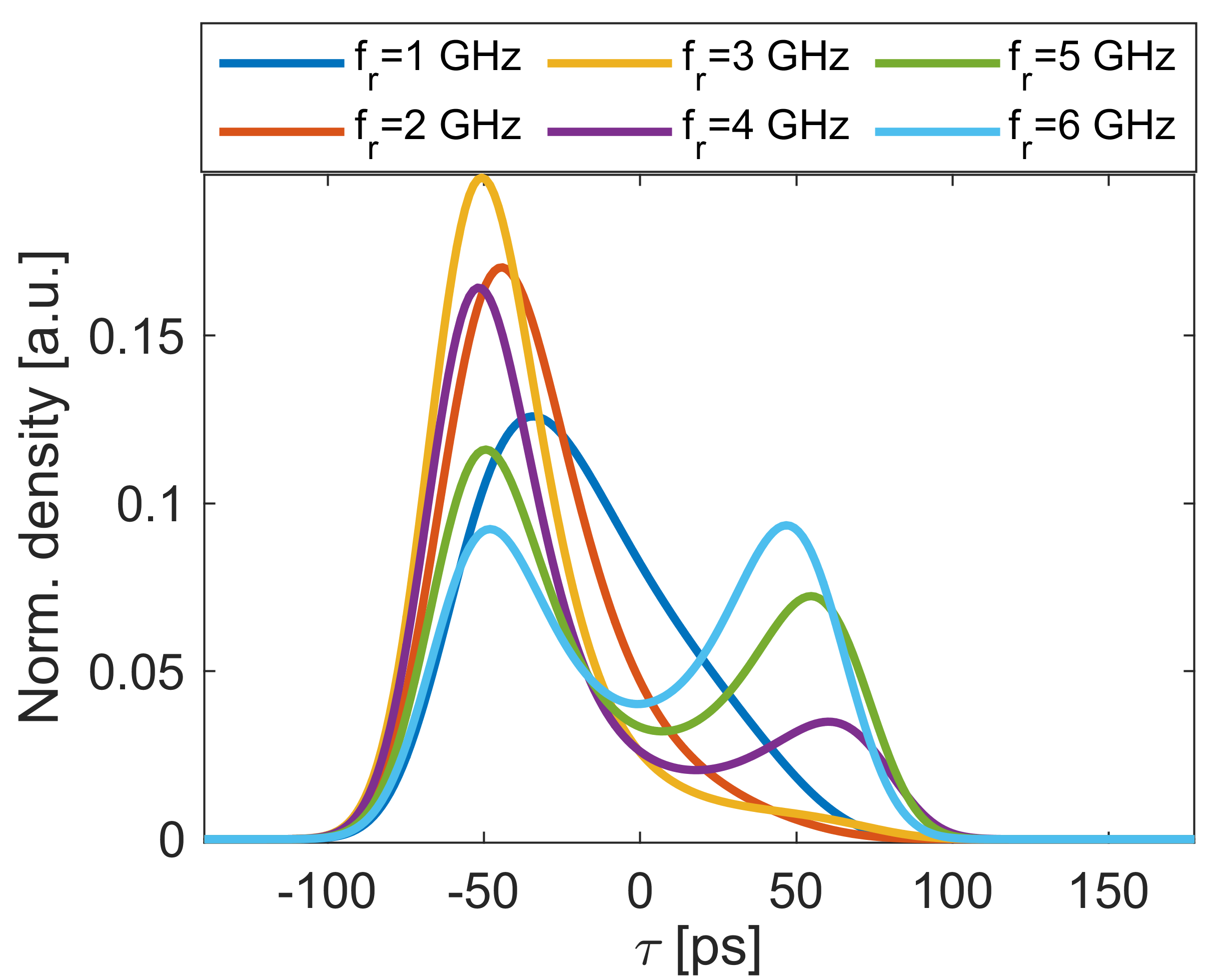}
  \caption{Bunch profiles for the case of assuming a high bunch charge of 20 nC. The legend gives the resonant frequency.}
  \label{fig9}
\end{figure}

It can be seen that the bunch lengthening can be degraded or enhanced by the short-range effects, which depends strongly on the resonant frequency. For HALF, it can be concluded that the short-range effects of resonant modes below 3 GHz will deteriorate the bunch lengthening; while for the case of higher than 4 GHz, it can enhance the bunch lengthening and even form a double-hump bunch profile, as shown in Fig.~\ref{fig9}.

For resonant modes in RF cavities, the fundamental mode typically exhibits the largest $R/Q$ value, while the $R/Q$ values of parasitic modes generally decrease with increasing resonant frequency. In synchrotron light sources, RF cavities predominantly show significant resonant modes below 3 GHz. Consequently, the short-range effects of RF cavity resonant modes are expected to consistently degrade bunch lengthening performance.

To demonstrate the significant influence of resonant mode short-range effects on bunch lengthening, we analyze a hypothetical high-bunch-charge scenario for HALF. This scenario enables us to isolate and evaluate how these short-range effects impact bunch lengthening. Notably, the HALF storage ring remains unaffected by these short-range effects under normal operating conditions. However, for synchrotron light sources employing multiple RF cavities with large $R/Q$ values—especially those operating in timing mode with high bunch charge—it is essential to account for these short-range effects. This consideration is critical when optimizing bunch lengthening performance using HC. 

\section{\label{sec:level4} Application to the PETRA-IV storage ring}
In Table~\ref{table3}, we summarize the information on bunch charge and the total $R/Q$ of main and harmonic cavities for the 4th-generation high-energy synchrotron light sources, including HEPS~\cite{Zhang26}, APS-U~\cite{APSU09}, ESRF-EBS~\cite{Jacob27}, and PETRA-IV~\cite{Ebert28}. Considering the short-range effects contributed by both MCs and HCs, we find that PETRA-IV exhibits the strongest short-range effects. Consequently, its short-range effects are expected to have the most significant impact on bunch lengthening.

\begin{table*}[!hbt]
\setlength{\tabcolsep}{6.5mm}
   \centering
   \caption{Bunch charge and cavity parameters of four high-energy synchrotron light sources.}
   \begin{tabular}{lccc}
       \toprule
       \textbf{Light source} &{Timing-mode} &{MC total $R/Q$}  &{HC total $R/Q$ } \\
         & bunch charge & @ Frequency & @ Frequency \\
       \hline
           HEPS & 14.4 nC& 348 $\Omega$ @ 166.6 MHz & 95 $\Omega$ @ 499.8 MHz \\
           APS-U & 15.3 nC   & 1248 $\Omega$ @ 352 MHz & 52 $\Omega$ @ 1408 MHz \\
           ESRF-EBS & 28 nC  & 943 $\Omega$ @ 352 MHz & 356 $\Omega$ @ 1409 MHz \\
           PETRA-IV& 7.68 nC & 2757 $\Omega$ @ 500 MHz&2117 $\Omega$@ 1500MHz\\
       \hline
   \end{tabular}
   \label{table3}
\end{table*} 

\subsection{\label{sec:level4.1} Bunch lengthening under the optimum-lengthening setting}
\begin{figure}[!htbp]  
  \centering
      \includegraphics[width=0.36\textwidth]{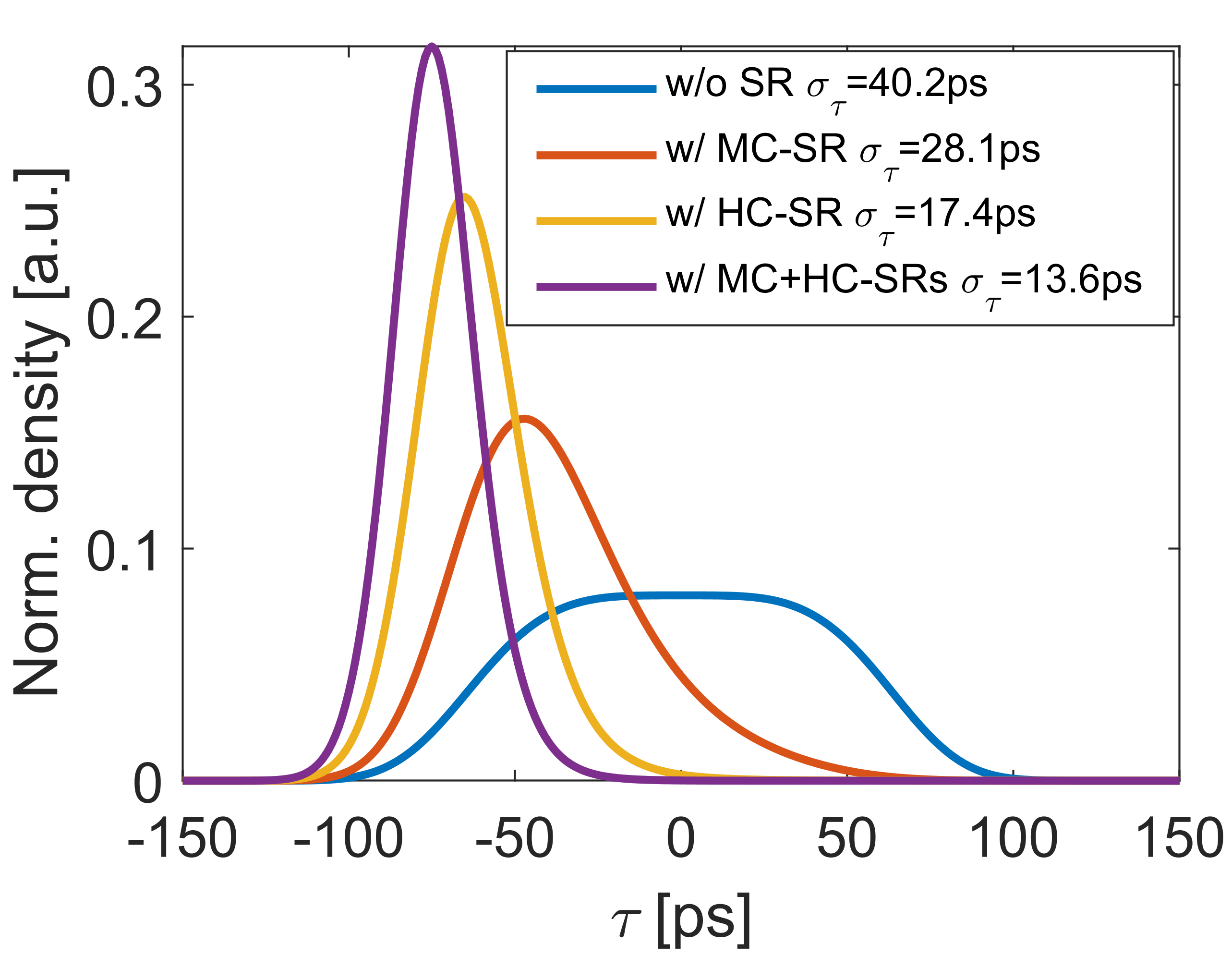}
  \caption{Bunch profiles for PETRA-IV in the case of flat-potential setting and maximum bunch charge required for the timing-mode operation.}
  \label{fig10}
\end{figure}

Figure~\ref{fig10} shows the resulting bunch equilibrium distributions for PETRA-IV under the optimum-lengthening setting of the double RF cavities, calculated using parameters from Table~\ref{table3} and Ref.~\cite{Agapov29}. Four cases abbreviated as the legends shown in Fig.~\ref{fig10} are considered, same as those of HALF shown in Fig.~\ref{fig2}. We can see that without the short-range effects, the optimum-lengthening RMS bunch length is 40.2 ps. However, the inclusion of the short-range effects of the double RF cavities can reduce the bunch length down to 13.6 ps, only one-third of the above optimum one. The reduction in bunch lengthening is more significant in the case of "w/ HC-SR" compared with the case of "w/ MC-SR", although the HC total $R/Q$ is slightly lower than that of the MC. This phenomenon is attributable to the HC's resonant frequency being triple that of the MC, enhancing the short-range effects.
\begin{figure*}[!hbt]  
  \centering
      \includegraphics[width=0.90\textwidth]{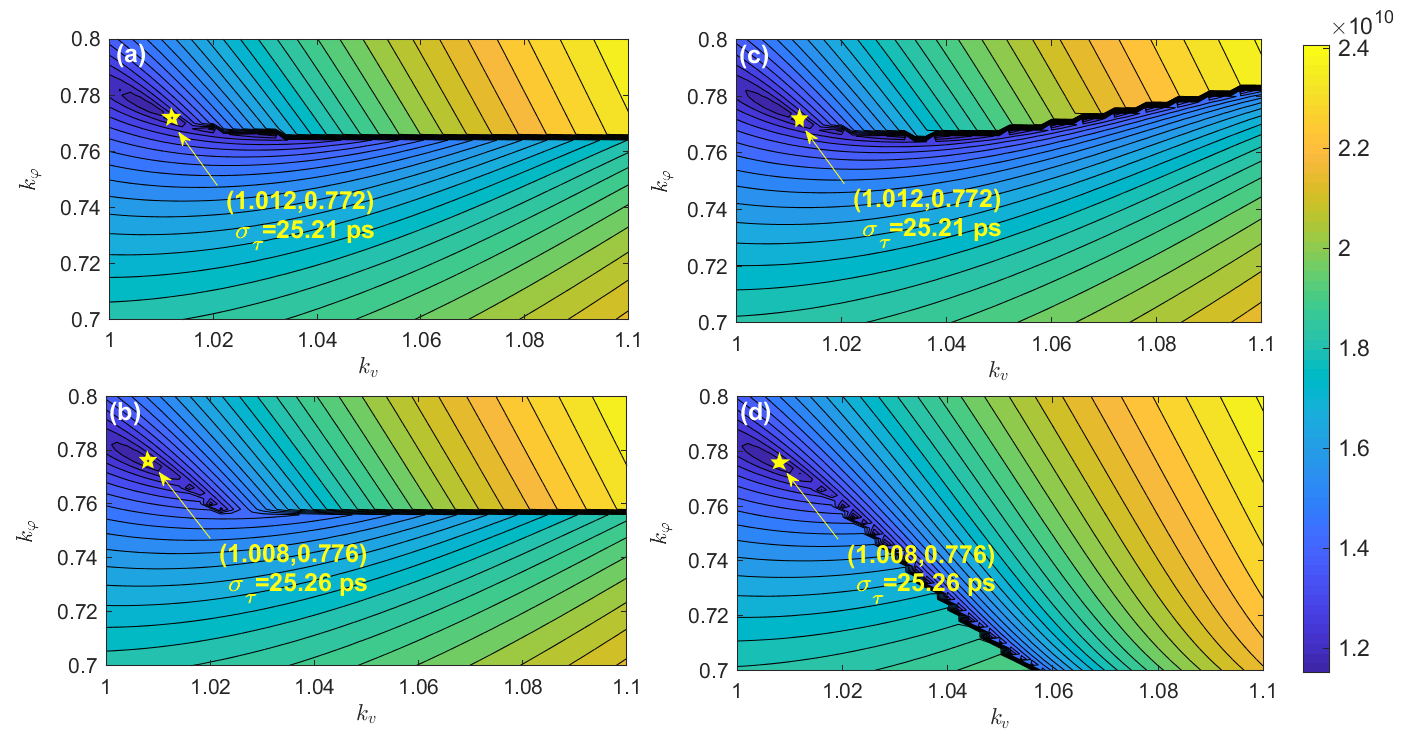}
  \caption{The contour map of the scanning solution of the integral $\int_{-\infty}^{\infty} \rho^2(\tau)\,d\tau$. (a) scanning $k_v$ from 1 to 1.1, (b) scanning $k_v$ from 1.1 to 1, (c) scanning $k_\varphi$ from 0.7 to 0.8, (d) scanning $k_\varphi$ from 0.8 to 0.7. The scanning step is set to 0.002. The yellow asterisk corresponds to the optimum solution.}
  \label{fig11}
\end{figure*}
\subsection{\label{sec:level4.2} Optimization of bunch lengthening}
It is a natural idea to optimize bunch lengthening by adjusting the HC voltage and phase. This can be achieved through systematic scanning of HC voltage and phase in the two-dimensional parameter space to identify the optimal bunch lengthening solution. To facilitate this scanning, we introduce two coefficients of $k_v$ and $k_{\varphi}$, and define that
\begin{equation}\label{eq:labe25}
\left\{
\begin{aligned}
&k = k_vk_{fp}  \\
&\varphi_n = k_{\varphi}\varphi_{n,fp}
\end{aligned}
\right.,
\end{equation}
where $k_{fp}$ and $\varphi_{n,fp}$ respectively denote the ratio of HC voltage to MC voltage and synchronous phase under the flat-potential condition. This condition is satisfied when $k_v=1$ and $k_{\varphi}=1$. Assuming that the optimized setting is close to the flat-potential setting, we can scan $k_v$ and $k_{\varphi}$ within a range around 1.

It should be noted that when optimizing the bunch lengthening, it is not seeking the maximum of bunch lengthening, but rather the maximum of Touschek lifetime. The improvement ratio of Touschek lifetime is characterized by~\cite{Byrd02}:
\begin{equation}\label{eq:labe26}
  \mathcal{R}=\frac{\int_{-\infty}^{\infty} \rho_0^2(\tau)\,d\tau}{\int_{-\infty}^{\infty} \rho^2(\tau)\,d\tau}.
\end{equation}
where $\rho(\tau)$ is the longitudinal density distribution of the stretched bunch, and $\rho_0(\tau)$ is the zero-current longitudinal bunch density distribution in the absence of HC, indicating that the numerator in Eq.~(\ref{eq:labe26}) can be regarded as a real constant. Therefore, the optimization objective should be to minimize the denominator in Eq.~(\ref{eq:labe26}).

We conducted two rounds of parameter scanning. In the first round, using a larger step size of 0.005 across broader ranges ($k_v$: 0.8-1.2, $k_\varphi$: 0.6-1.1), the optimal solution was identified at $k_v=1.015$ and $k_\varphi=0.77$. For the second round, based on initial findings, we narrowed the scanning ranges to $k_v$: 1.0-1.1 and $k_\varphi$: 0.7-0.8 with a finer step size of 0.002. Four distinct scanning methods were implemented: fixing either $k_v$ or $k_\varphi$ while systematically scanning the other parameter in both descending and ascending orders. Figure~\ref{fig11} shows the scanning results, where the yellow asterisk indicates the optimum solution. 

First, we observe that the optimum solutions obtained from the four scanning methods are twofold: "$k_v=1.012, k_\varphi=0.772$" and "$k_v=1.008, k_\varphi=0.776$". Although the numerical values are slightly different, they are very close to each other, and the corresponding bunch length is approximately 25 ps. Compared with the unoptimized case, the bunch length has been improved by about a factor of two. Secondly, we find that the contour distributions of the solutions from the four scanning methods are the same in some regions but different in others, indicating the existence of double-valued solutions in specific areas.

\begin{figure}[!htbp]  
  \centering
      \includegraphics[width=0.45\textwidth]{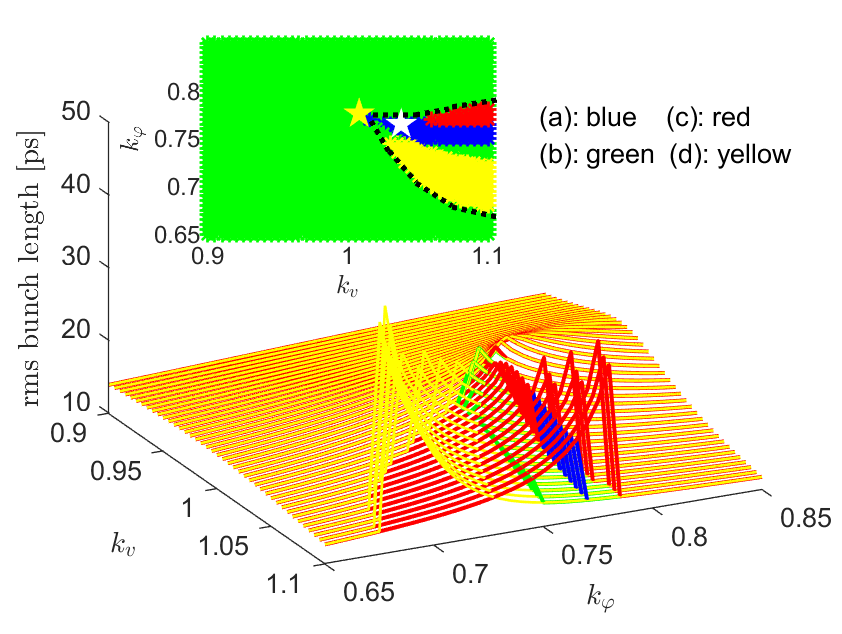}
  \caption{The 3D plot of the scanning solution of rms bunch length as functions of $k_{v}$ and $k_{\varphi}$. The legends denote the four scanning cases as same as those shown in Fig.~\ref{fig11}. In the subplot, there exist two equilibrium solutions within the area enclosed by the black-dashed lines, and the yellow and white stars correspond to the settings of "$k_v=1.008,k_\varphi=0.776$" and "$k_v=1.038,k_\varphi=0.765$", respectively.}
  \label{fig12}
\end{figure}

To obtain more detailed information, we have plotted the three-dimensional (3D) distributions of the resulting bunch length, as shown in Fig.~\ref{fig12}. The regions with double-valued solutions are enclosed by black-dashed lines, as indicated in the subplot of Fig.~\ref{fig12}. In addition, we have selected two specific parameter settings: "$k_v=1.008, k_\varphi=0.776$", and "$k_v=1.038, k_\varphi=0.765$". The former corresponds to the optimum solution shown in Fig.~\ref{fig11} (c) and (d) and is represented by a yellow asterisk, while the latter is located within the region enclosed by the black-dashed lines and corresponds to the double-valued solutions, represented by a white asterisk. It can be seen that the optimum solution is very close to the region with double-valued solutions.

\begin{figure}[!htbp]  
  \centering
      \includegraphics[width=0.40\textwidth]{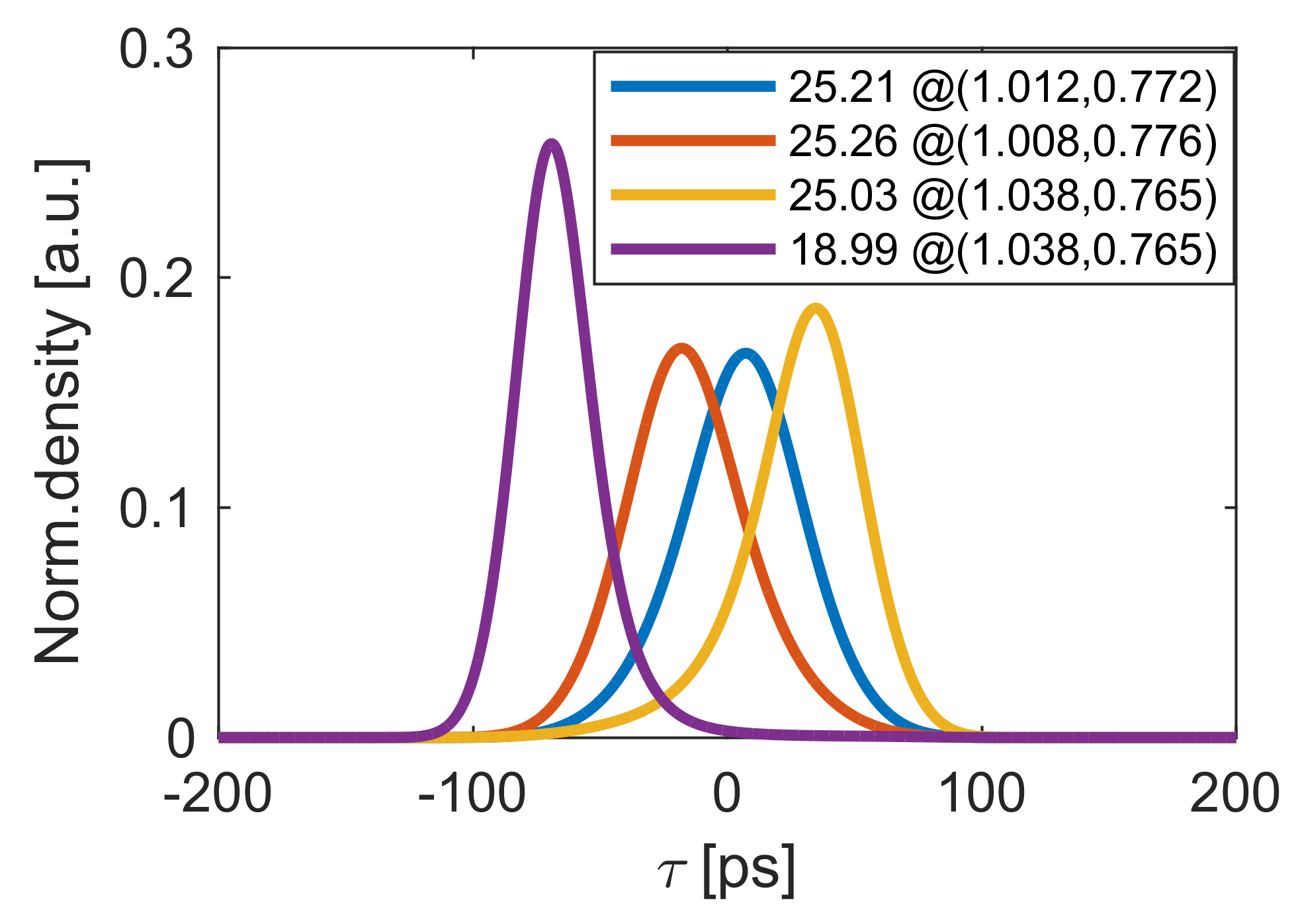}
  \caption{Bunch profiles. The legend gives the corresponding rms bunch length in units of ps and the setting of ($k_v$, $k_\varphi$).}
  \label{fig13}
\end{figure}

Figure~\ref{fig13} shows the bunch profiles corresponding to the two optimum solutions as illustrated in Fig.~\ref{fig11}, where the solutions are unique. Additionally, it also displays the bunch profiles corresponding to "$k_v=1.038, k_\varphi=0.765$", where two solutions exist: one with the centroid shifted forward and a bunch length of approximately 19 ps, and the other shifted backward with a bunch length of about 25 ps. We have verified the existence of these double-valued solutions using tracking simulation methods, and more details can be found in Ref.~\cite{He31}.

In summary, for PETRA-IV, we have demonstrated the results of the scanning optimization, proving that there exists an optimum HC parameter setting that can achieve nearly a factor of two improvement in bunch length compared to the theoretical optimum setting. We have also shown that this optimum setting is very close to the domain of double-valued solutions.

It should be noted that the above optimization can in general work for an active HC, whose voltage amplitude and phase can be flexibly set. Another important point to note in the above optimization is that only the short-range effects from RF cavity fundamental modes are taken into account. When it is necessary to consider parasitic modes, or the total broadband impedance, or both, we can still follow the above optimization idea that optimizing HC voltage and phase to maximize the Touschek lifetime improvement ratio.

\section{\label{sec:level5} Conclusion}
In this paper, we investigated the short-range effects of RF cavity resonant modes on bunch lengthening in high-bunch-charge scenarios. We found that:
(i) Short-range effects have a minor impact on bunch lengthening when without HC operation, but they significantly influence lengthening when HC is set for optimum lengthening.
(ii) In the optimum HC setting, the short-range effects of fundamental modes in double RF cavities can substantially reduce bunch lengthening. Parasitic modes have short-range effects independent of $Q$, which can also significantly affect bunch lengthening if their $R/Q$ is non-negligible.
(iii) For timing-mode operation with high bunch charge and multiple RF cavities with large total $R/Q$, the short-range effects of all resonant modes must be considered to accurately evaluate the beam equilibrium distribution.
(iv) When resonant modes have strong short-range effects, the conventional optimum lengthening setting is insufficient, and alternative HC parameters are needed to maximize the Touschek lifetime. (v) Under specific HC settings, there exist two equilibrium bunch distributions.  These findings are crucial for synchrotron light sources using multiple RF cavities and aiming for optimum lengthening in high-bunch-charge scenarios.

\begin{acknowledgments}
This research was supported by the National Natural Science Foundation of China (No. 12375324 and No. 12105284) and the Fundamental Research Funds for the Central Universities (No. WK2310000127).
\end{acknowledgments}

\appendix
\renewcommand{\appendixname}{APPENDIX~}
\section{\label{sec:AppdenxA}TRACKING SIMULATIONS FOR HALF WITH A PASSIVE HC UNDER NEAR-OPTIMUM LENGTHENING CONDITION}
This paper focused on the optimum lengthening condition, which can be generally satisfied when using active HCs. As a supplement, we would like to show similar results obtained when using passive HCs that detuned to achieve the near-optimum bunch lengthening. We still used the HALF storage ring as an example to demonstrate the impact of the short-range effect of beam-fundamental mode interaction on bunch lengthening. The STABLE-code was used to conduct the tracking simulations~\cite{HeGPU30}. Recently, a Proportional-Integral feedback module has been added to STABLE to stabilize the main cavity voltage, which can be used to simulate the low-level RF feedback effects. The passive HC is detuned to 55 kHz to meet the near-optimum bunch lengthening condition at 350 mA. Keeping the beam current of 350 mA constant, the number of uniformly filled bunches is gradually reduced from 800 to 4. In this way, the high bunch charge scenario can be created artificially, while the averaged HC voltage amplitude can be kept nearly unchanged. Each bunch is represented by 50,000 macro-particles and 50,000 turns are tracked. We conducted two rounds of calculations, considering and ignoring the short-range effect of beam-fundamental mode interaction.

\begin{figure}[!htbp]
   \centering
   \includegraphics*[scale=0.53]{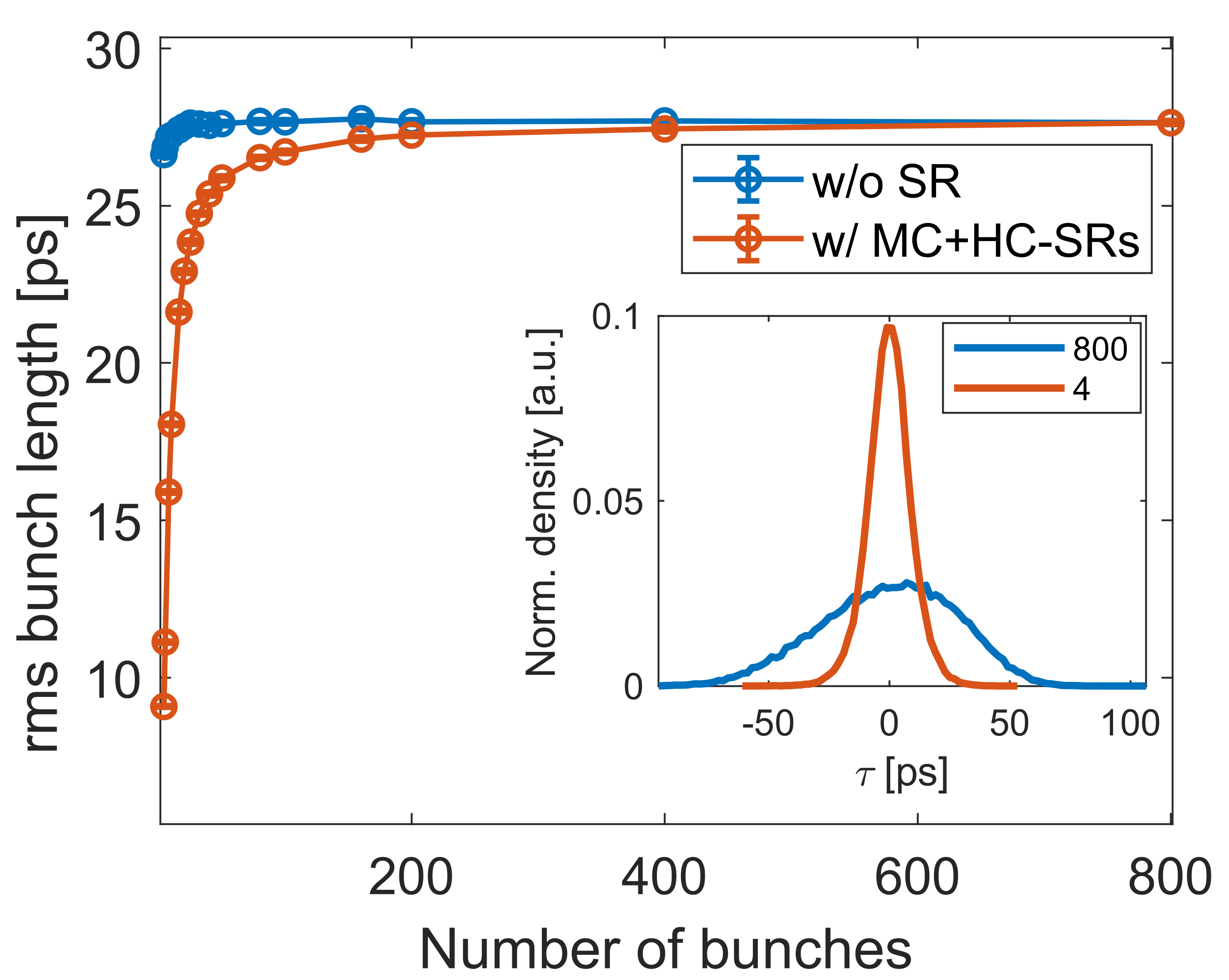}
   \caption{The rms bunch length vs. the number of filled bunches obtained with and without the short-range effects. The number of filled bunches are set to 800, 400, 200, 160, 100, 80, 50, 40, 32, 25, 20, 10, 8, 5, 4, respectively. Bunch profiles in the cases of 800 bunches filling and 4 bunches filling are shown in the subplot.Only one bunch is shown due to the uniformly filling setting.}
   \label{fig14}
\end{figure}

The resulting bunch lengths versus the number of filled bunches are shown in Fig.~\ref{fig14}. The subplot shows the bunch profiles in the case of 800 and 4 filled bunches.
It is evident that the short-range effects can significantly reduce bunch lengthening, particularly when the number of filled bunches is 4. This underscores the importance of considering the strong short-range interaction between the beam and fundamental modes to accurately evaluate the bunch equilibrium.

\nocite{*}

\end{document}